\documentclass{aa}
\usepackage{txfonts}
\usepackage{graphicx}

\def\tvi(#1,#2){\vrule height #1pt depth #2pt width 0pt}
\def\p{\partial}
\def\e{{\rm e}}
\def\d{{\rm d}}
\def\ie{i.e. }
\def\eg{e.g. }
\def\etal{et al. }

\def\M{{\cal M}}

\def\omcutu{\omega_1^{\rm cut}}
\def\omcutz{\omega_0^{\rm cut}}
\def\rsh{{r}_{\rm sh}}

\def\g{\Theta}

\begin{document}

\title{A fresh look at the unstable simulations of  
Bondi-Hoyle-Lyttleton accretion }
\author{T. Foglizzo\inst{1}, P. Galletti\inst{1}, \and M. Ruffert\inst{2}}
\offprints{foglizzo@cea.fr}

\institute {Service d'Astrophysique, CEA/DSM/DAPNIA, CEA-Saclay, 91191Gif-sur-Yvette, France\and School of Mathematics, University of Edinburgh, Edinburgh EH9 3JZ, Scotland}

\date{Received 18 October 2004 / Accepted 2 February 2005}

\abstract{
The instability of Bondi-Hoyle-Lyttleton accretion, observed in  
numerical simulations, is analyzed
through known physical mechanisms and possible numerical artefacts. The  
mechanisms of the longitudinal and transverse instabilities,  
established within the accretion line model, are clarified. They cannot  
account for the instability of BHL accretion at moderate Mach number  
when the pressure forces within the shock cone are taken into account.
The advective-acoustic instability is considered in the context of BHL  
accretion when the shock is detached from the accretor. This mechanism  
naturally explains the stability of the flow when the shock is weak,  
and the instability when the accretor is small. In particular, it is a  
robust proof of the instability of 3D accretion when $\gamma=5/3$ if  
the accretor is small enough, even for moderate shock strength ($\M\sim  
3$).
The numerical artefacts that may be present in existing numerical  
simulations are reviewed, with particular attention paid to the  
advection of entropy/vorticity perturbations and the artificial  
acoustic feedback from the accretor boundary condition. Several  
numerical tests are proposed to test these mechanisms.
}

\titlerunning{Instability mechanism}
\maketitle

\section{Introduction}

The phenomena described by numerical simulations are usually highly  
simplified compared to the physical reality. However simplified, these  
phenomena are sometimes complicated enough to challenge our ability to  
understand them. The flow of gas onto a gravitating accretor moving  
supersonically is a classic astrophysical problem (Hoyle \& Lyttleton  
1939, Bondi \& Hoyle 1944) which is relevant in many astrophysical  
contexts such as wind fed X-ray binaries, supermassive black holes,  
star formation, and also galaxies in a cluster (see the recent review  
by Edgar 2004).
Early numerical simulations in 2D (Matsuda \etal 1987, Fryxell \& Taam  
1988) revealed that this flow is unstable. 3D simulations (Matsuda  
\etal 1991, Ruffert \& Arnett 1994) confirmed the unstable character of  
the flow, displaying however a weaker variability than in 2D.
The instability is not understood even in the simple case of an ideal  
uniform gas.
Is this instability physical, or numerical? Is it the same instability  
mechanism in 2D and 3D? How can this unstable behaviour be extrapolated  
to small accretor sizes, which are relevant for wind fed X-ray binaries  
but out of reach of numerical simulations?
Some authors recently doubted that this instability is physical  
(Pogorelov, Ohsugi \& Matsuda 2000). The most recent relativistic  
simulations by Font \& Ibanez (1998a, 1998b) and Font, Ibanez \&  
Papadopoulos (1999) also showed stable flows.
What about the published instability mechanisms which have been  
proposed over the years? What is left of the longitudinal and  
transverse  instabilities of the accretion line (Cowie 1977, Soker  
1990, 1991), the "relatively simple" mechanism based on the shock  
opening angle by Livio \etal (1991), the "vortex shedding in the Von  
Karman manner" (Koide \etal 1991, Matsuda \etal 1991, 1992)?
The present uncertain situation reveals how unconvincing the proposed  
mechanisms were. Some were inconclusive, lacked quantitative criteria,  
and others might also have been incompletely understood. The present  
work aims at clarifying the different instability mechanisms and  
confront them with existing numerical simulations, paying particular  
attention to possible numerical artefacts.\\
The general trends based on the existing simulations, and the proposed  
mechanisms are recalled in Sect.~2. The longitudinal and transverse  
instabilities of the accretion line are revisited in Sect.~3 and  
Sect.~4. The advective-acoustic instability is adapted to the BHL flow  
in Sect. ~5.
This enables a new look at the simulations in Sect.~6, in an attempt to  
reconcile them.
The basis of new simulations, free of numerical artefacts, testing  
these ideas, is described in Sect.~7. For the sake of the clarity of  
the paper, the main text contains only the most important equations,  
which summarize the analytical arguments proven in appendices A to H.

\section{An overview of existing simulations of BHL accretion and  
proposed instability mechanisms}

\subsection{Numerical simulations are numerous}

\begin{table*}
\caption[]{Overview of the published numerical simulations of BHL  
accretion using a polytropic equation of state and a totally absorbing  
accretor: plane accretion in 2D, 3D axisymmetric accretion and full 3D  
accretion. The first column contains an abreviated reference which can  
be found in the bibliography.}
\label{tableoverview}
\centering
\begin{tabular}{ccccccc}
\hline\hline
          ref.           & grid   &     Mach    &   accretor  &    index  
    &    transverse   & stability \\
                       &             &   number  &   $r_*/r_{\rm A}$  
&$\gamma$ & gradients& \\
\hline\\
3D& & & & & & \\
\hline\\
R99     &    Cart.  &  1.4 - 10 &   0.02 - 1  &   1.01, ${4/3}$,  
${5/3}$ &  den & no \\
R97     &    Cart.  &  0.6 - 10 &   0.02 - 1  &   ${4/3}$, ${5/3}$    &  
    vel & no \\
R96       &  Cart.    & 0.6 - 10  &  0.02 - 1   &  ${1.01}$   & -  &     
no     \\
R95       &  Cart.    & 0.6 - 10  &  0.02 - 1   &  ${4/3}$   & -  &     
no    \\
RA95 &  Cart.    &  3      &   0.01 - 10  &  ${5/3}$    & vel &        
no \\
RA94 &  Cart.    &  3      &   0.01 - 10  &  ${5/3}$    & - &       no   
\\
R94       &  Cart.    & 0.6 - 10  & 0.02 - 10    & ${5/3}$     & -  &   
no    \\
IMS93  &   cyl.  & 3      &     0.125    &   isothermal   &   den/vel   
& no\\
MIS92 &  Cart.     & 3                  &   0.1   &  $1.005$ - ${5/3}$   
   & - &       no           \\
MSS91  & Cart.   &  3          & 0.06 - 0.25    & ${5/3}$                
& - &   no  \\
SMA89 & curv. & 1.4& 0.1 & ${5/3}$& vel& quasi \\
LSK86       & Cart.   &   3, 16   &  0.15   &   ${7/6}$ - ${5/3}$    &  
den & quasi  \\
SLK86        & Cart.   &   2, 4   &  0.15   &   $1$    & den & yes  \\
\hline\\
3D axisym.& & & & & & \\
\hline\\
POM00 & polar   &    3 - 10  &  0.05    &    1.01, 1.4, ${5/3}$   & - &  
        yes   \\
FI98a &   polar &   0.6 - 10  &   0.1 - 2.4 &     1.1, ${4/3}$, ${5/3}$  
   & - & yes   \\
KMS91      &  polar      &  1.4 - 10  & 0.005 - 0.015  & ${5/3}$         
& - &     no  \\
MSS89   &   polar &  1.4    &   0.01 - 0.05  &     ${5/3}$      & - &    
no       \\
SMA89 & curv. & 1.4& 0.1 & ${5/3}$&  - &yes \\
PSS89   & polar & 0.6 - 5   &  0.125  & 1.1, ${4/3}$, ${5/3}$, 2   & -  
&  yes \\
FTM87 & polar & 1.4 -  4 & 0.016 - 0.13& ${5/ 3}$ &- & no\\
SMT85&  polar & 0.6 - 5 &0.1 &1.1, ${4/3}$, ${5/3}$& -&yes  \\
H79 &  polar & 0.6 - 3.6&0.01 & ${4/3}$& -& - \\
H71 &  polar &0.6 - 2.4 &0.01&${5/3}$ & -& -\\
\hline\\
2D planar& & & & & & \\
\hline\\
POM00 & polar     &    3 - 10  &  0.05    &    1.4, ${5/3}$   &  
-/den/vel&        yes  \\
POM00 & polar     &    3 - 10  & 0.05     &    1.01   & -&         
no/yes   \\
POM00 & polar     &    4  &   0.05  &    ${4/3}$   & den&        no      
\\
FIP99  & polar   &   5      &  0.25  &    ${4/3}$, ${5/3}$, 2         &  
- &    yes   \\
FI98b &  polar & 3 - 10 &  0.25    &      1.1, ${4/3}$, ${5/3}$     & -  
& yes  \\
SMA98   & polar    &     1 - 16   & 0.005 - 0.05 & isothermal     & - &  
   no  \\
BLT97 &   polar &  4    &    0.001, 0.005   &   ${4/3}$    & -/den/vel  
&     no \\
ZWN95 & Cart.& 3 &0.03 - 0.13 & ${5/3}$ & - & yes \\
ZWN95 & Cart.& 4 &0.03 - 0.13 & ${4/3}$ & den & no \\
BA94 & SPH & 3&0.04 - 0.13 &1.1, 1.3, 1.5 & - & no  \\
IMS93  &   Cart.  &   3      &     0.13   &    isothermal   &  den/vel   
& no \\
MIS92 &  Cart.    & 1.4 - 10    &     0.04 - 0.3    &   1.005 - ${5/3}$  
  & - &        no          \\
MSS91 &  Cart.  &   3 - 5    &  0.06 - 0.25  &  1.2, ${5/3}$    & - &    
no     \\
SMA89 & curv. & 3& 0.1& 1.5, 2& vel&no  \\
TF89 &  polar &    4     &    0.037   &     ${4/3}$       &  vel  &no    
\\
FT88 &  polar &    4     &    0.037   &     ${4/3}$       &  den  &no    
\\
MIS87  & curv.  &  1 - 5   &   0.03 - 0.6  &    ${4/3}$, ${5/3}$    &   
binary & no \\
ABM87        & SPH   &   3    & 0.13, 0.15   &   1.5    & vel, den &  
yes\\
\hline
\end{tabular}
\end{table*}

Numerical simulations of the BHL problem started with the work of Hunt  
(1971). The instability first appeared in Matsuda \etal (1987), Fryxell  
\& Taam (1988). The many subsequent simulations are listed in  
Table~\ref{tableoverview}. Simulations involving more complicated  
ingredients such as realistic heating and cooling (\eg Blondin \etal  
1990, Taam, Fu \& Fryxell 1991) are not included for the sake of  
simplicity. The simulations of  Table~\ref{tableoverview} are divided  
into three groups corresponding to plane accretion, axisymmetric  
accretion and full 3D accretion.
In each of these groups, the listing in chronological order follows the  
progress in computing speed and numerical techniques over the last 30  
years. This progress enabled the simulation of smaller and smaller  
accretors, improving the first attempts by a factor 10. Denoting by  
$r_{\rm A}\equiv 2GM/v_\infty^2$ the accretion radius of an accretor of  
mass $M$ and velocity $v_\infty$, the most recent simulations reach  
$r_*/r_{\rm A}=0.005$ in axisymmetric flows (KMS91), $r_*/r_{\rm  
A}=0.001$ in planar accretion (BLT97), and $r_*/r_{\rm A}=0.01$ in 3D  
accretion (RA94). Global trends can be summarized as follows:
\par-The shock is always attached to the accretor in simulations of 2D  
planar flow. By contrast, 3D simulations revealed a detached bow shock,  
ahead of the accretor, if the accretor is small enough and  
$\gamma\ge4/3$. A calculation in Appendix~A suggests that the shock  
should be detached in planar flows with $\gamma\sim3$.
\par-Although the strength of the instability varies from one code to  
another, the instability is found in numerical simulations using any of  
the coordinate systems, Cartesian, polar, cylindrical or special  
curvilinear, even with SPH. If numerical, the phenomenon is not  
specific to a particular grid or a specific method.
\par-Simulations of plane accretion exhibit the most unstable  
behaviour. The shock moves sideways in a flip-flop manner (MIS87). The  
instability is strongest for small accretors, possibly for intermediate  
Mach numbers ($\M=3$ according to MIS87 and POM00).
\par-The presence of velocity or density gradients in the transverse  
direction of the flow was considered in the earliest unstable  
simulations (MIS87, FT88, SMA89), but MSS91 realized that this  
ingredient is not crucial for instability. This conclusion was  
challenged by ZWN95 who confirmed the flip-flop instability when the  
accretor is a square single cell (as in MSS91), but found a stable flow  
when the accretor is spatially resolved and modelized as a polygon.  
 From their point of view, the instability is related to the detachement  
of the shock.
\par -Stable planar accretion flows were also found by FI98b and FIP99  
who considered a rather large ratio $r_*/r_{\rm A}=0.25$ and POM00  
whose method is further discussed below.
\par-Axisymmetric flows are generally stable, with few exceptions.  
Among them, SMT85, FTM87 and MSS89 showed vortex shedding when the  
accretor is a hard, non absorbing sphere : in this case the problem is  
similar to the classical flow around a  sphere, modified by gravity. An  
important exception to the stability of axisymmetric flow is KMS91, who  
also found vortex shedding for an absorbing accretor, if $\M\ge2.4$. It  
can be noted that the accretor size they considered is the smallest  
ever used in axisymmetric simulations.
\par-The instability of full 3D accretion is never as strong as the  
flip-flop observed for planar flows, but seems present in all published  
simulations, at least for small enough accretors.
Even the earliest simulations of LSK86 and SMA89 showed some persistant  
oscillations. The instability seems to be more violent if the shock is  
detached ($\gamma=4/3$ and $5/3$).

\subsection{Several mechanisms were proposed}

The question of the stability of BHL accretion could in principle be  
solved by performing a perturbation analysis on a stationary solution  
such as obtained by POM00. This procedure would be very heavy,
and has never been achieved. Using various simplifications, six  
different physical instability mechanisms have been proposed so far. In  
chronological order:
\par (i) longitudinal instability of the accretion line (Cowie 1977),
\par (ii) transverse instability of the accretion line (Soker 1990,  
1991),
\par (iii) shock opening angle (Livio \etal 1991),
\par (iv) vortex shedding in the Von Karman manner (Koide \etal 1991,  
Matsuda \etal 1991, 1992),
\par (v)  local Rayleigh-Taylor (RT) and Kelvin-Helmhotz (KH)  
instabilities (Foglizzo \& Ruffert 1999),
\par (vi) advective-acoustic cycle (Foglizzo \& Tagger 2000, Foglizzo  
2001, 2002).\\
Some of the proposed mechanisms (iii, iv, v) are interesting ideas,  
which need to be developed, but which are not conclusive at present:
\par -The calculation of Livio \etal (1991) is not a stability analysis  
of the shock surface, but rather an attempt to express in equations the  
idea that the pressure should decrease along the shock surface.
Since no growth rate or typical timescale is computed, extrapolating on  
the possibility that this is responsible for the violent and chaotic  
behaviour observed in 2D simulations is not convincing.
\par -Vortex shedding, by the interaction of the incoming gas with the  
"atmosphere" captured by the accretor, was demonstrated by SMT85 and  
MSS89 in simulations where the accretor is non-absorbing.
The fate of this instability in BHL accretion, where the gas is  
absorbed at supersonic velocity, is rather speculative.
\par -The local analysis of FR99 investigates two natural causes of  
instability (RT and KH), and concluded that these are not  
quantitatively convincing without a feedback mechanism.\\

Among the cited mechanisms, the only conclusive stability analysis are  
those of Cowie (1977) and Soker (1990, 1991) in the approximation of  
the accretion line model. This simplification is known to be valid only  
when the shock opening angle is very narrow, \ie at very high Mach  
number. Even then, it misses the possible interaction between a bow  
shock and the accretor (as described by the advective-acoustic cycle).  
However simplified, these instability mechanisms are physical. They  
should be understood well enough to predict what is left of them beyond  
the accretion line model.

\section{A new look at the longitudinal instability of the accretion  
line}

\subsection{Physical cause of the longitudinal instability}

The mechanism of the longitudinal instability of the accretion line was  
briefly explained by Cowie (1977) as being due to the effect of  
accreted momentum on density perturbations. Soker (1990) challenged  
this explanation by assessing that this instability mechanism is  
independent of accretion and is a mere consequence of the acceleration  
of the flow. A closer look at the equations,  in Appendix~B, shows the  
weakness of this argument. Let us consider more generally the following  
dynamical system:
\begin{eqnarray}
{\p \rho\over \p t}+{\p \rho v\over \p r}&=&H(r),\label{dzH}\\
{\p v\over \p t}+v{\p v\over \p r}&=&F(r,v,\rho),\label{dyF}
\end{eqnarray}
where $H,F$ are regular functions describing the mass input and the  
force per unit mass. The particular case of the accretion line model is  
described in Table~\ref{tableacline} and Appendix~A, where velocities  
are in units of $v_\infty$, distances $r$ are in units of the accretion  
radius $r_{\rm A}$ , and the line density $\rho$ is normalized using  
the mass accretion rate (the normalization of distances and densities  
used by Cowie (1977) are different by a factor $2$).
\begin{table}
\caption[]{Accretion line models in 2D and 3D. }
\label{tableacline}
\begin{tabular}{ccc}
                  & 2D & 3D\\
\hline
\\
$H$     &    $\displaystyle{r^{-{1\over2}}}$ &  $1$\\
\\
$F$     &    $\displaystyle{{1-v\over r^{1\over2}\rho}-{1\over 2r^2}}$   
&
$\displaystyle{{1-v\over \rho}-{1\over 2r^2}}$\\
\\
\hline
\end{tabular}
\end{table}
A linearization of Eqs.~(\ref{dzH}-\ref{dyF}) gives the differential  
equation satisfied by a perturbation of the mass flux $h\equiv  
\rho_0\delta v+v_0\delta\rho$ (Appendix~B).
In what follows, the subscript for unperturbed quantities $v_0,\rho_0$  
is omitted.
The solution is written at high frequency $\omega$ using the WKB  
approximation  :
\begin{eqnarray}
h\sim\left({\rho\over v {\p F\over\p\rho}}\right)^{1\over4}
\exp\left\lbrack\int\left( v{\p F\over\p v}-\rho{\p  
F\over\p\rho}\right){\d r\over 2v^2}
\right\rbrack\nonumber\\
\exp\left\lbrack i\omega\int{\d r\over v}
\pm {1-i\over 2^{1\over2}}\omega^{1\over 2}\int \left({\rho\over  
v^3}{\p F\over\p \rho}\right)^{1\over2}
\d r\right\rbrack.\label{wkbh}
\end{eqnarray}
The flow is thus unstable at high frequency if the force per unit mass  
$F$, acting on the accretion line, depends on density. The instability  
does not depend on the accretion of mass, as stressed by Soker (1990),  
in the sense that it does not depend on the function $H$. Nevertheless  
it does depend on the accreted momentum through the function $F$. In  
this sense, accretion plays a crucial role in this instability, as  
initially sketched by Cowie (1977). Contrary to the conclusions of  
Soker (1990), acceleration within the accretion line is not crucial for  
this instability (see a counter example in Appendix~B).
In the 3D accretion line model, Eq.~(\ref{wkbh}) becomes:
\begin{eqnarray}
h(r)\propto {r^{1\over2}\over\log^{1\over4}r}\exp\left\lbrack i\omega r
\pm{1+i\over 3}(2\omega)^{1\over2}\log^{3\over2}  
r\right\rbrack,\nonumber\\
{\rm for}\;\;r\gg\alpha,\label{ampliout3}\\
h(r)\propto r^{5\over8}\exp
\left\lbrack-{2\over3}i\omega r^{3\over 2}
\pm{1-i\over 5}\left({8\omega\over\alpha}\right)^{1\over2}
r^{5\over4}\right\rbrack,\nonumber\\
{\rm for}\;\;{1\over\omega^{2\over 5}}\ll r\ll\alpha,\label{ampliin3}
\end{eqnarray}
where $\alpha$ is the distance of the stagnation point. The amplitude  
of high frequency perturbations increases far from the accretor  
($r\gg\alpha$).
Perturbations in the accreting part of the flow ($r<\alpha$) are also  
unstable at high frequency $\omega$ down to a point $r\propto  
\omega^{-{2\over5}}$ where the amplication ceases and the WKB  
approximation breaks down. In a numerical simulation, the treatment of  
high frequency perturbations is limited by the numerical resolution.  
According to Eqs.~(\ref{ampliout3}-\ref{ampliin3}), the better the  
resolution, the stronger the instability on both sides of the  
stagnation point. Formulae computed in Appendix~B for 2D flows show  
slight differences which are not significant on the scale of a few  
accretion radii.

\subsection{The longitudinal instability modified by pressure forces:  
analogy with radiation driven winds \label{radwind}}

\begin{figure}
\begin{center}
\includegraphics[width=\columnwidth]{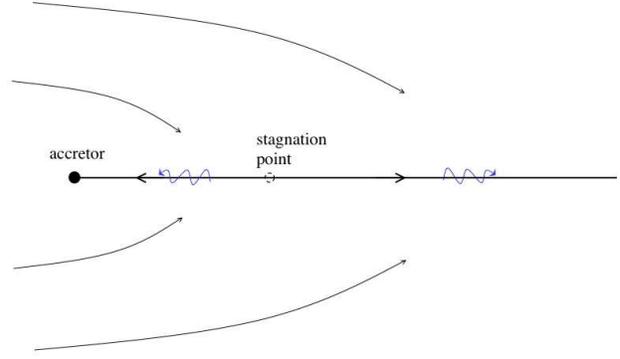}
\caption[]{Schematic view of the accretion line model. Stream lines are  
drawn as solid lines with arrows. When pressure is taken into account  
(Sect.~\ref{radwind}), density perturbations in the accretion line  
propagate as acoustic waves (wavy lines with arrows).}
\label{figaccline}
\end{center}
\end{figure}

The region of the stagnation point is necessarily subsonic, and  
longitudinal pressure forces cannot be neglected there. Pressure forces  
in the stationary accretion line were considered by Wolfson (1977),  
Yabushita (1978a) and Horedt (2000). The simplest formulation  
corresponds to the isothermal hypothesis, in which the dynamical  
equations (\ref{dzH}-\ref{dyF}) are changed into:
\begin{eqnarray}
{\p \rho\over \p t}+{\p \rho v\over \p r}&=&H(r),\\
{\p v\over \p t}+v{\p v\over \p r}&=&F(r,\rho,v)-{c^2\over \rho}{\p  
\rho\over\p r},
\end{eqnarray}
where $c$ is the isothermal sound speed.
Again, a single differential equation is obtained for the radial  
structure of the linear perturbation $h$ of the mass flux  
(Eq.~(\ref{singlediffpress})),
describing the propagation of acoustic waves modified by the external  
forces $F$. The solution is approximated at high frequency through a  
WKB analysis, away from the sonic points ($v=\pm c$):
\begin{eqnarray}
h\sim\rho^{1\over2}
\exp\int {\d r\over v\pm c}\left\lbrack i\omega+{1\over2}\left(
{\p F\over\p v}
\pm{\rho\over c}{\p F\over\p \rho} \right)\right\rbrack,\nonumber\\
{\rm for}\;\;\omega\gg {\rho v\over c^2}{\p  
F\over\p\rho}.\label{freqwkb}
\end{eqnarray}
This stability analysis resembles that of radiation driven winds,  
studied by Mestel, Moore \& Perry (1976) and Mathews (1976).  
Considering a uniform gravitational acceleration $g$ and a linearized  
radiative force ($F\equiv A\rho-g$, $H\equiv 0$), they showed that  
acoustic waves propagating outwards are amplified by radiation.
In addition to the density effect $\p F/\p\rho$, the velocity effect  
$\p F/\p v$ is formally well known in line-driven winds (see \eg  
Carlberg 1980, and more recent reviews by Owocki 1994, Feldmeier \&  
Owocki 1998). $\p F/\p v <0$ in the accretion line model, implying that  
this velocity effect is always stabilizing, independently of the  
direction of propagation. This could have been anticipated directly  
from the Euler equation, since a positive perturbation of velocity  
results in a decreased external force.
By contrast, the effect of the density dependence (term $\p F/\p\rho$)  
is opposite for outgoing and ingoing waves:
\begin{eqnarray}
{\p F\over\p v}\pm{\rho\over c}{\p F\over\p \rho}
&=&-{v\over r-\alpha}\left(1\pm {v_\infty-v\over c}\right)\;\;{\rm  
in\;3D},\label{accline}\\
&=&-{v\over r^{1\over 2}(r^{1\over 2}-\alpha)}\left(1\pm  
{v_\infty-v\over c}\right)\;\;{\rm in\;2D}.
\label{accline2D}
\end{eqnarray}
The density effect $\p F/\p\rho<0$ is thus stabilizing for waves  
propagating outwards and destabilizing for waves propagating inwards.  
This can be understood as follows: with $\p F/\p\rho<0$, a positive  
density perturbation is associated with a decreased external force.  
According to the Euler equation, this decreased force has a damping  
effect on the positive velocity perturbation associated with a wave  
propagating outwards, whereas it amplifies the negative velocity  
perturbation associated with a wave propagating inwards. In contrast  
with the instability without pressure found by Cowie (1977), this  
possible amplification of ingoing acoustic waves is not oscillatory.  
Altogether, the density and velocity effects damp outgoing acoustic  
waves.
According to Eqs.~(\ref{freqwkb}) and (\ref{accline}-\ref{accline2D}),  
acoustic waves propagating inwards may be amplified only if
\begin{eqnarray}
c+v<v_\infty.
\end{eqnarray}
This condition cannot be fulfilled far from the accretor since $v\sim  
v_\infty$ for $r\gg\alpha$. From this we conclude that the only  
possible amplification of high frequency acoustic waves is restricted  
to ingoing waves in a region of finite size. The size of this region is  
independent of frequency. The amplification factor ${\cal A}$, deduced  
from the WKB analysis,
is also independent of the perturbation as long as its frequency is  
high enough to satisfy Eq.~(\ref{freqwkb}). It can be estimated as  
follows in 3D:
\begin{eqnarray}
{\cal A}&\sim& \exp\int {v_\infty-v-c\over 2c}{v\over r-\alpha}{\d  
r\over v-c}
\;\;{\rm for}\;\;\omega\gg{v^2\over c^2}{v_\infty-v\over  
r-\alpha}.\label{ampliA}\\
&\le&\e^{\M_{\infty}}
\end{eqnarray}
This contrasts with the instability found by Cowie (1977) which could  
be arbitrarily fast at high frequency  
(Eqs.~\ref{ampliout3}-\ref{ampliin3}).

\subsection{The longitudinal instability beyond the accretion line  
model}

The longitudinal instability of the accretion line proves that if the  
Mach number is high enough, the amplification of ingoing acoustic waves  
should be visible in numerical simulations.
Such a transient amplification, however, cannot be considered to be a  
convincing mechanism to explain the instability observed in numerical  
simulations for moderate Mach numbers $\M_\infty\sim 3-5$.
A true instability would require a feedback loop, in order to build an  
acoustic cycle. Ingoing acoustic waves may be partially reflected  
outwards near the accretor. Outgoing waves, however, are likely to  
escape to infinity rather than be reflected inwards again. Moreover,  
their amplitude is damped by the effect of the accreted momentum. In  
conclusion, the longitudinal instability of the accretion line does not  
explain the instability of BHL accretion.

\section{A new look at the transverse instability of the accretion line}

\subsection{High frequency approximation of the transverse instability}

Soker (1990) extended the stability analysis of Cowie (1977) to the  
case of transverse perturbations in 2D planar flows.
The position of the accretion line is described in polar coordinates  
$\theta=\g(r)$. The angle $\Psi$ between the tangent to the accretion  
line and the symmetry axis, and the transverse velocity $v_\theta$ are  
related to $\g$ as follows:
  \begin{eqnarray}
\Psi=\g+r{\p \g\over\p r},\\
v_\theta = r\left({\p \over\p t}+v{\p\over\p r}\right)\g.
\end{eqnarray}
As remarked by Soker (1990), the transverse instability is decoupled  
from the longitudinal one. The angle $\g$ of the accretion line  
satisfies a second order differential equation, which is approximated  
at high frequency in Appendix~D using a WKB analysis:
\begin{eqnarray}
\g&\propto& {1\over r^{5\over4}}\exp \left\lbrack i\omega r
  \pm 2^{1\over2}(1+i)\omega^{1\over 2}r^{1\over2}\right\rbrack\;\;{\rm  
for}\;\;r\gg\alpha,\label{gsup}\\
\g&\propto& {1\over r^{7\over8}}\exp\left\lbrack i\omega\int{\d r \over  
v}
\pm {2(1-i)\over5\alpha^{1\over2}}\omega^{1\over 2}  
r^{5\over4}\right\rbrack
\;\;{\rm for}\;\;r\ll\alpha.\label{ginf}
\end{eqnarray}
According to Eq.~(\ref{gsup}), the larger the distance from the  
accretor, the larger the amplification of perturbations. As for the  
longitudinal instability, the amplification in the region of accretion  
ceases close to the accretor at $r\propto \omega^{-2/ 5}$, and thus  
depends on the numerical resolution.
These high frequency estimates could be used to compare the radial  
profiles of the transverse and longitudinal instabilities of the  
accretion line in the linear regime in 2D flows: the differences  
between Eqs.~(\ref{gsup}-\ref{ginf}) and  
Eqs.~(\ref{ampliout2}-\ref{ampliin2}) are not significant on the scale  
of few accretion radii, and cannot be responsible for the dominant  
longitudinal instability observed in the non linear simulations of  
Soker (1991).

\subsection{Comparison of the transverse instability with the  
instability of a flag}

More generally, a system satisfying a transverse equation of the form
\begin{eqnarray}
\left({\p \over\p t}+v{\p\over\p  
r}\right)v_{\theta}=A\g+B\Psi+Cv_{\theta},\label{geneflop}
\end{eqnarray}
with arbitrary coefficients $A,B,C$, is unstable at high frequency. The  
differential equation (\ref{diflag}) satisfied by $\g$ is written in  
Appendix~D. If $v\ne0$, the WKB approximation at high frequency is as  
follows:
\begin{eqnarray}
\g\propto \left({v\over r^2B}\right)^{1\over4}  
\exp\int\left\lbrack{i\omega \over v}
+{C\over2v}+{B\over2v^2}\pm\left({i\omega B\over v^3}\right)^{1\over 2}
\right\rbrack\d r,\nonumber\\
{\rm for}\;\;\omega\gg {v\over r},C,{Av\over rB}, {\p\log\over\p  
r}{B\over v^2}.\label{geneflop2}
\end{eqnarray}
This formulation outlines the role of the restoring force $B \Psi$ in  
driving the high frequency instability. $B\equiv -1/(r^{1\over2}\rho)$  
in the 2D accretion line model. This mechanism is reminiscent of the  
instability of a flag as described by Argentina et al. (2004), where  
the hydrodynamical force acting on the flag is also proportional to the  
inclination $\Psi$.
The equation describing the transverse motion of a flag with infinite  
flexibility corresponds to the same Eq.~(\ref{geneflop}), with $v\equiv  
0$, $A\equiv0$, $B\equiv -\alpha U_0$ and $C\equiv\alpha$, where $U_0$  
is the wind velocity and the coefficient $\alpha>0$ characterizes the  
aerodynamic force acting on the flag. The solution of  
Eq.~(\ref{geneflop}) when $v\equiv 0$ is unstable at high frequency  if  
$B<0$:
\begin{eqnarray}
\g\propto {1\over r} \exp\int \left\lbrack
i\omega {C\over B}-{\omega^2\over B}-{A\over rB}\right\rbrack\d r.
\end{eqnarray}
The presence of finite flexibility ("flexural rigidity") in a realistic  
flag material would set an upper bound to unstable frequencies. The  
comparison with the instability of a flag suggests the possible  
existence of a transverse instability of the subsonic region of the  
stagnation point ($v\sim 0$).

\subsection{The transverse instability beyond the accretion line model}

\begin{figure}
\begin{center}
\includegraphics[width=\columnwidth]{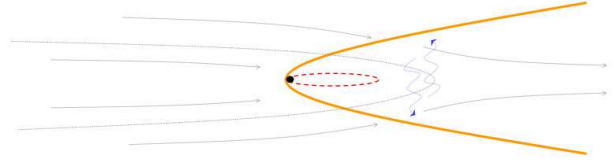}
\caption[]{Transverse instability growing in the vicinity of the  
stagnation point. The shock (thick line), the sonic surface (dashed  
line) and the stream lines (thin lines with arrows) are drawn. The thin  
dotted line delineates the accreted gas. The transverse motion of the  
shock on one side is transmitted to the other side through acoustic  
waves (wavy lines with arrows).
}
\label{figflag}
\end{center}
\end{figure}

\subsubsection{Transient growth of the transverse instability in a  
shock cone}

The accretion line model relies on the asumption that the half angle  
$\theta_0$ of the shock cone is small compared to other distances.  
Identifying the shock cone with the Mach cone,
$\theta_0\sim1/\M_\infty$ is indeed small if the incident flow is  
highly supersonic.
With a longitudinal wavelength comparable to $\sim 2\pi v/\omega$, the  
transverse instability of the accretion line model is valid only for
\begin{eqnarray}
{2\pi v\over\omega}\gg{2 r\over\M_{\infty}}.
\end{eqnarray}
The maximum exponential amplification ${\cal A}$ deduced from  
Eqs.~(\ref{gsup}-\ref{ginf}) is thus bounded by:
\begin{eqnarray}
{\cal A}\le \exp (2\pi\M_{\infty})^{1\over2}.
\end{eqnarray}
Taking into account the finite width of the shock cone sets an upper  
bound $\propto \M_\infty$ to the frequency of the most unstable  
perturbations in a plane 2D flow. An additional limitation to the  
accretion line model comes from the acoustic time across the shock  
cone, which is a lower bound to the growth time of the instability.  
This also favours the subsonic region surrounding the stagnation point  
rather than the supersonic regions away from it.
The instability as described by Soker (1990) is a local mechanism,  
taking place in the advected flow. A global mode as observed in the 2D  
plane simulations of the flip-flop instability requires an acoustic  
feedback inside the subsonic region.
This instability could be described as a purely acoustic cycle between  
the opposite sides of the shock cone as in Fig.~\ref{figflag}.

\subsubsection{Towards a global transverse instability in BHL accretion  
?}

The amplification of transverse perturbations should be visible if the  
Mach number is high enough. Whether this is sufficient to explain the  
instability observed in 2D flows at low Mach number is not clear since  
this instability is transient.
When pressure is taken into account, the region of the stagnation point  
seems to be a privileged place for the growth of a global mode  
involving transverse displacement and acoustic propagation. \\
The fate of the transverse instability in 3D BHL accretion is rather  
uncertain, as remarked by Soker (1991). Obviously transverse motions of  
the accretion line are forbidden in 3D, since incoming symmetrical  
trajectories which do not intersect the displaced accretion line meet  
along the symmetry axis and generate a new accretion line. This does  
not exclude a possible unstable oscillation of the shock cone in a  
transverse direction, such that the symmetry axis stays inside the  
accretion shock. According to Eqs.~(\ref{geneflop}-\ref{geneflop2}),  
the existence of a transverse restoring force proportional to the  
inclination angle is enough to generate a high frequency instability.  
Beyond geometrical factors, this force is still present in the 3D  
geometry. This instability mechanism could thus be present for high  
enough Mach numbers, although it has never been observed clearly in 3D  
numerical simulations. A correct description of the stability with  
respect to transverse oscillations would require taking into account  
pressure effects and the 3D dynamical deformation of the shock cone,  
which is much more difficult than the accretion line formalism.

\section{The advective-acoustic mechanism in BHL accretion\label{adac}}

\subsection{Schematic formulation of a global cycle}

The advective-acoustic instability deals with the cycle of advected  
perturbations (entropy, vorticity) coupled to acoustic waves, between  
the shock and the sonic surfaces. The coupling at the shock is a local  
process associated to the conservation laws through the shock. By  
contrast, the coupling due to the inhomogeneity of the subsonic flow  
occurs all the way from the shock to the sonic surface. F01 showed in a  
simple radial geometry that this acoustic feedback is described by an  
integral over the subsonic region, dominated by the region close to the  
sonic point, where the temperature is highest. This allows us to  
decompose the advective-acoustic cycle in four steps as follows:
\par(1) advection of an entropy/vorticity perturbation from the shock  
to the sonic point,
\par(2) excitation of an acoustic feedback due to the inhomogeneity of  
the flow,
\par(3) propagation of this acoustic feedback towards the shock,
\par(4) excitation of a new entropy/vorticity perturbation on the shock  
surface.\\
Each of these steps $j=1$ to $4$ is characterized by an efficiency  
${\cal Q}_j$ measuring the amplification of perturbations. This  
decomposition is motivated by the existence of invariants, allowing a  
direct calculation of ${\cal Q}_1$ based on the conservation of  
entropy, or ${\cal Q}_3$ based on the conservation of acoustic energy.
The stability of the global cycle depends on the product ${\cal Q}$:
\begin{eqnarray}
{\cal Q}\equiv{\cal Q}_1\times{\cal Q}_2\times{\cal Q}_3\times{\cal Q}_4
\end{eqnarray}
The linear growth rate of the advective-acoustic cycle, measured by the  
imaginary part of the eigenfrequency $\omega\equiv  
(\omega_r,\omega_i)$, can then be approximated by
\begin{eqnarray}
\omega_i\sim{1\over\tau}\log|{\cal Q}|,
\end{eqnarray}
where $\tau$ is the duration of the advective-acoustic cycle, generally  
dominated by the advection time. Note that this schematized approach  
neglects the purely acoustic cycle, which can influence the stability  
threshold (FT00, F02).

\subsection{Efficiencies ${\cal Q}_i$ based on radial accretion}

In a radial flow with $\gamma>1$, we choose to measure, in the  
entropic-acoustic cycle, the amplification of the perturbation $f$ of  
the Bernoulli constant :
\begin{eqnarray}
f\equiv v\delta v+{2\over\gamma-1}c\delta c.
\end{eqnarray}
Let us denote by $f^\pm$ its values for pressure perturbations $\delta  
p^\pm$ corresponding to an energy flux $F^\pm$ propagating with $(+)$  
or against ($-$) the stream, and $f^S$ for entropy perturbations  
$\delta S$. At high frequency, for low degree waves $l=0,1,2$:
\begin{eqnarray}
F^\pm&\propto&{1\over\M c^2}|f^\pm|^2,\label{acflux}\\
f^\pm&\sim& (1\pm\M)c^2{\delta p^{\pm}\over\gamma p} ,\label{fpm}\\
f_S&\sim& {c^2\over\gamma}\delta S.\label{fS}
\end{eqnarray}
Non radial entropy perturbations $\delta S$ are associated to vorticity  
perturbations $\delta w$ in shocked spherical accretion through  
Eqs.~(\ref{wr}) to (\ref{wp}),
so that the amplification of the entropy $\delta S\rightarrow\delta  
p^-\rightarrow\delta S'$ also measures the simultaneous amplification  
of vorticity $\delta w\rightarrow\delta p^-\rightarrow \delta w'$. In  
the isothermal limit ($\gamma\rightarrow 1$), vorticity is more  
appropriate than entropy to describe the advective-acoustic cycle,  
which becomes a vortical-acoustic cycle (F02) .

The steps (1) and (3) of advection and propagation in the spherical  
accretion flow studied by F01, F02 are deduced from Eqs.~(\ref{acflux})  
and (\ref{fS}) and the
conservations of entropy $\delta S$ and acoustic energy $F^-$:
\begin{eqnarray}
{\cal Q}_1&\equiv& {f_{\rm son}^S\over f_{\rm sh}^S}\sim {c_{\rm  
son}^2\over c_{\rm sh}^2}
\;\; {\rm for}\;\;\omega_r\sim {c_{\rm son}\over r_{\rm  
son}},\label{Q1}\\
{\cal Q}_3&\equiv& {f_{\rm sh}^-\over f_{\rm son}^-}\sim \M_{\rm  
sh}^{1\over 2}{c_{\rm sh}\over c_{\rm son}},
\end{eqnarray}
where the subscripts "son" and "sh" refer to the sonic point and the  
shock respectively.
Although the acoustic energy is conserved, the amplitude of outgoing  
waves $f^-$ decreases (${\cal Q}_3<1$) due to the geometric dilution in  
a diverging flow.\\
The advection of an entropy perturbation in a hot region may greatly  
increase the thermal energy carried by this perturbation, by a factor  
comparable to the temperature ratio (Eq.~(\ref{Q1})). As sketched by  
FT00, the difference of energy is carried away by acoustic waves,
propagating both upward and downward. Even the waves propagating  
downward may be
refracted upward at low enough frequency (F01). The ratio ${f_{\rm  
son}^-/ f_{\rm son}^S}$ is thus of order unity:
\begin{eqnarray}
{\cal Q}_{2}\equiv {f_{\rm son}^-\over f_{\rm son}^S}\sim 1.
\end{eqnarray}
The advective-acoustic coupling ${\cal Q}_4$ at the shock is deduced  
from a local analysis of a perturbed shock, performed in Appendix~F:
\begin{eqnarray}
{\cal Q}_{4}\equiv {f_{\rm sh}^S\over f_{\rm sh}^-}\propto{1-\M_{\rm  
sh}\over\M_{\rm sh}}.\label{Q4}
\end{eqnarray}
This equation is identical to Eqs.~(16) and (C.11) in FT00, obtained  
for radial perturbations. ${\cal Q}_4$ may reach large values in nearly  
isothermal flows with strong shocks ($\M_{\rm sh}\sim 1/\M_1$), in  
which case the dominant advected perturbations are non radial (F02):  
this is the basis of the vortical-acoustic cycle. The factor  
$(1-\M_{\rm sh})$ in Eq.~(\ref{Q4}) is responsible for the damping of  
the advective-acoustic coupling for weak shocks. \\
Altogether,
\begin{eqnarray}
{\cal Q}\propto {c_{\rm son}\over c_{\rm sh}}{1-\M_{\rm sh}\over\M_{\rm  
sh}^{1\over2}}.
\label{qglobal}
\end{eqnarray}
This approximate formula illustrates the two regimes of efficient  
advective-acoustic coupling identified by FT00, F01, F02:
\par (i) Strong temperature gradients are responsible for an efficient  
triggering of acoustic waves from advected entropy perturbations, which  
is the basis of the entropic-acoustic cycle (FT00, F01).
\par (ii) Strong entropy/vorticity perturbations can be produced at the  
shock if the post shock Mach number $\M_{\rm sh}$ is small.\\
These two sources of amplification can be used as guidelines for  
anticipating the properties of advective-acoustic cycles in BHL  
accretion flows. However, the dependence of ${\cal Q}$ on the frequency  
and the degree $l$ of the perturbation requires further calculations.  
In particular, F01 showed that  the acoustic feedback is strongest for  
non radial modes $l=1$. Note also that Eq.~(\ref{qglobal}) is singular  
if $c_{\rm son}\to \infty$ as is the case for 3D accretion with  
$\gamma=5/3$, which deserves a more careful analysis (Appendix~G).

\subsection{From radial to BHL accretion\label{radialBHL}}

\begin{figure}
\begin{center}
\includegraphics[width=\columnwidth]{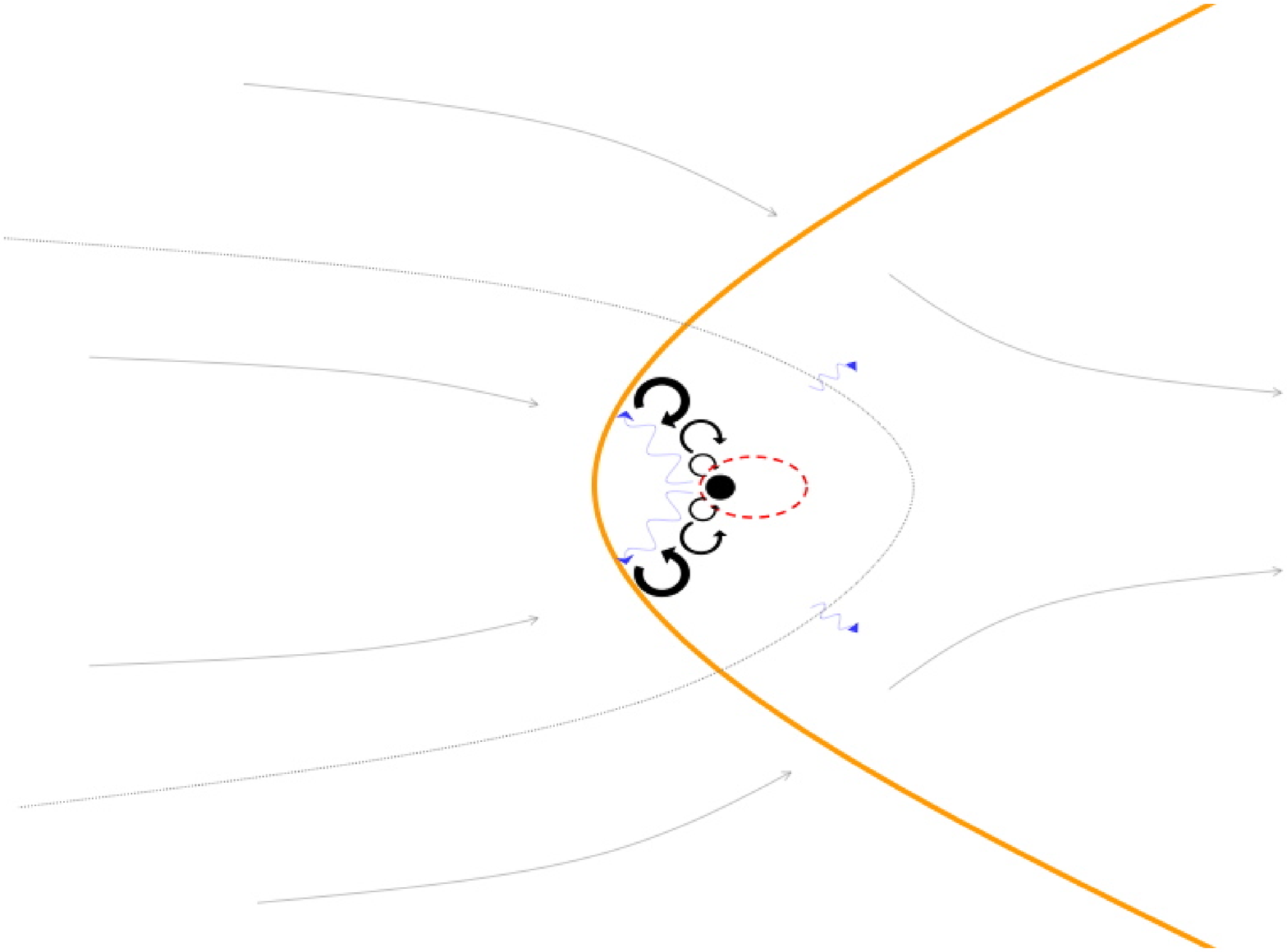}
\includegraphics[width=\columnwidth]{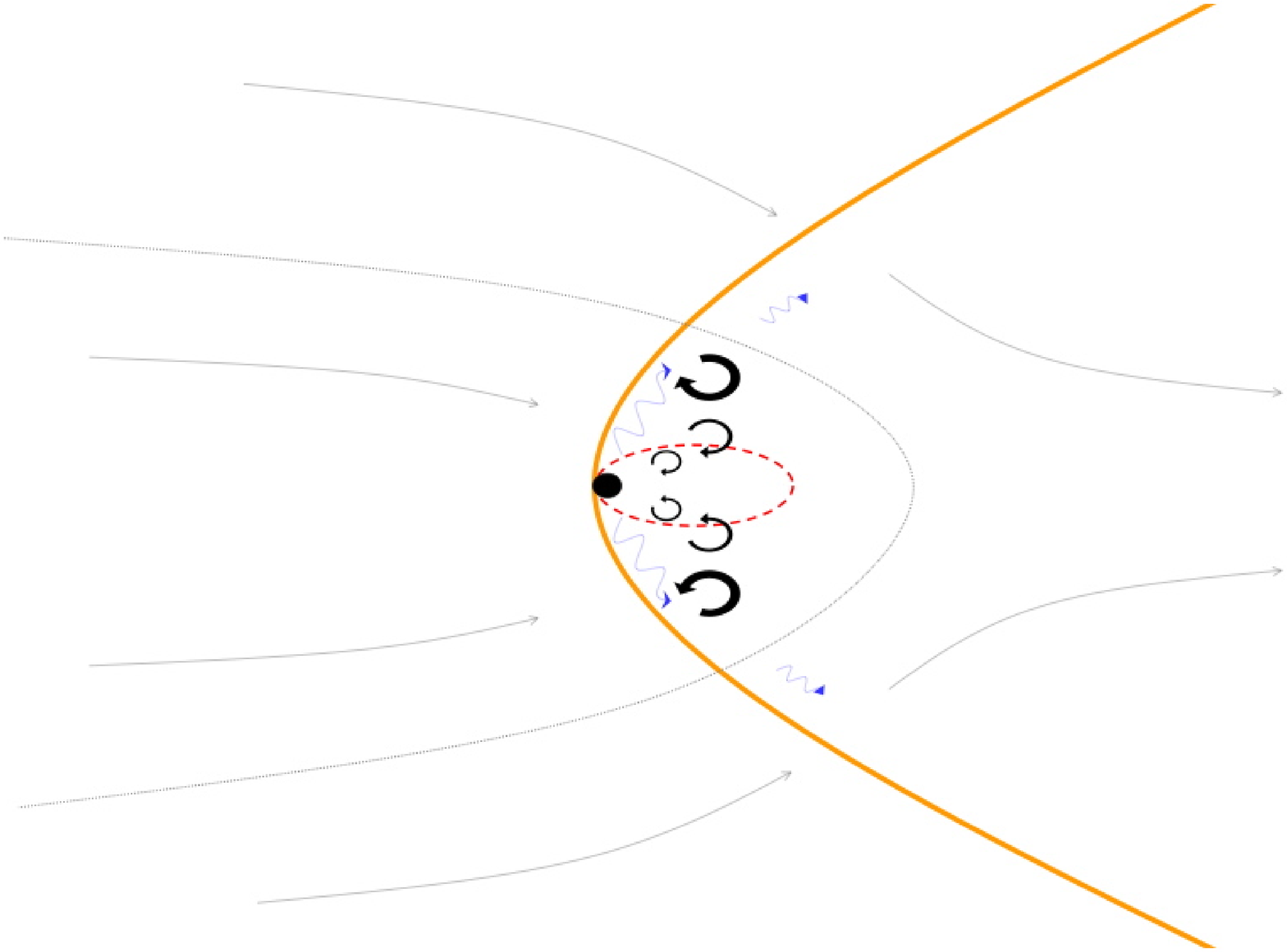}
\caption[]{Schematic view of a global mode if the shock is detached or  
attached. Drawn lines have the same meaning as in Fig.~\ref{figflag}.  
Perturbations of entropy/vorticity (circular arrows) follow the flow  
lines from the shock to the accretor, producing an acoustic feedback  
which propagates to the shock. A fraction of the acoustic energy may  
leak outside of the accretion cylinder. }
\label{figmode}
\end{center}
\end{figure}

\subsubsection{Extrapolation to detached shocks}

The locus of the advective-acoustic cycle is different if the shock is  
attached or detached (Fig.~\ref{figmode}). A global mode between a bow  
shock and the accretor resembles the advective-acoustic cycle in a  
shocked radial flow. In this case the efficiency of the  
advective-acoustic cycle can be estimated from the study of radial  
accretion.
By contrast, if the shock is attached to the accretor, most of the flow  
is accreted from behind in a supersonic manner. The value of ${\cal  
Q}_2$ in this particular geometry cannot be directlly extrapolated from  
its value in a radial flow.

\subsubsection{Geometrical factors}

In BHL accretion, the shape of the shock surface is not only  
non-spherical but open to infinity.
The value of ${\cal Q}_3$ might be reduced by a geometrical factor  
$\sim 2$ compared to spherical accretion because a significant fraction  
of acoustic waves may propagate away from the region of accretion and  
leave the cycle (Fig.~\ref{figmode}).\\
Conversely, the value of ${\cal Q}_1$ should be increased by the  
amplification of vorticity perturbations through the local KH and RT  
mechanisms (FR99), due to vorticity and entropy gradients in the  
post-shock flow. \\
The efficiencies ${\cal Q}_i$ estimated in spherical geometry should  
thus be considered, at best, as very rough approximations of those in  
the BHL flow.

\subsection{Influence of the parameters $\gamma$, $\M_\infty$ and $r_*$}

According to FT00, F01, the stability of the entropic-acoustic cycle  
depends essentially on the
temperature increase between the shock and the sonic point. The most  
unstable cycle involves high frequency acoustic waves, those able to  
explore the hottest parts of the flow but still be refracted out, with  
a wavelength slightly larger than the smallest size of the sonic  
surface. Using the Bernoulli equation, the temperature on a point of  
the sonic surface is directly related to its distance $r_{\rm son}$ to  
the accretor:
\begin{eqnarray}
c_{\rm son}^2= {\gamma-1\over\gamma+1}\left\lbrack
{2GM\over r_{\rm son}}+c_\infty^2\left(  
\M_\infty^2+{2\over\gamma-1}\right)\right\rbrack.
\end{eqnarray}
The closer the sonic surface to the accretor, the higher the  
temperature, and the more efficient the entropic-acoustic cycle.
The calculation of Appendix~A indicates that the sonic surface is  
always attached to the accretor if $\gamma=\gamma_{\rm max}$, with
\begin{eqnarray}
\gamma_{\rm max}&\equiv&3\;\;{\rm in\;2D},\\
\gamma_{\rm max}&\equiv&{5\over 3}\;\;{\rm in\;3D},
\end{eqnarray}
even if the shock is detached. Fig.~17 of R94 shows the shape of the  
attached sonic surface for 3D accretion with $\gamma=5/3$. Together  
with Eq.~(\ref{qglobal}), the strength $\cal Q$ of the instability  
should increase when $\gamma$ approaches $\gamma_{\rm max}$ and $r_*$  
decreases. The strength of the instability should be asymptotically  
independent of the incident Mach number for strong shocks. Conversely,  
the instability should be suppressed if the shock is weak ($\M_{\rm  
sh}\sim 1$). \\
According to Appendix~A, a critical index $\gamma_{\rm  
crit}<\gamma_{\rm max}$ must exist, below which the shock is always  
attached to the accretor, whatever its size. Numerical simulations with  
$\gamma=4/3$ (R95) suggest that $\gamma_{\rm crit}<4/3$ in 3D flows.
The entropic-acoustic cycle is expected to be an efficient instability  
mechanism in the range $[\gamma_{\rm crit},\gamma_{\rm max}]$, as long  
as the distance of the sonic surface is small enough. \\
Nearly isothermal flows ($\gamma\sim 1$) could be unstable through the  
vortical-acoustic cycle, fed by ${\cal Q}_4\gg1$ for strong shocks.  
However, the effect of the acoustic feedback in the particular geometry  
of an attached shock is uncertain. Neither the vortical-acoustic  
mechanism nor the extrapolated transverse instability manage to explain  
why isothermal BHL accretion seems so much more unstable in planar  
flows (SMA98) than in 3D simulations (R96).

\section{Can the different numerical simulations be reconciled?}

\subsection{Some simulations are stable\label{stable}}

The stability of the 2D planar flow simulated by FI98b, FI99, with a  
relativistic accretor contrasts with the many unstable simulations  
performed in Newtonian gravity.
This is by no mean a relativistic stabilization of BHL accretion  
through relativistic effect,
as recognized by FI98b.  Indeed,  they restricted their studies to  
$v_\infty/c_{\rm light}=0.5$, which corresponds to a rather big  
Schwarzschild radius in units of the accretion radius: $r_{\rm  
Sch}/r_{\rm A}=0.25$. These flows would have been stable even in the  
Newtonian limit.\\
According to ZWN95, the flow simulated by MSS91 becomes stable if the  
square accretor is replaced by a polygon. Although the shape of the  
accretor may influence the instability threshold,
the existence of strongly unstable simulations in polar coordinates  
(BLT97 and SMA98) shows that 2D accretion can be unstable even if the  
accretor is perfectly spherical. The smaller accretor size and higher  
resolution used by BLT97 and SMA98, compared to ZWN95 (see  
Table~\ref{tableoverview}), may be a hint in favour of an  
advective-acoustic mechanism.
A direct comparison, however, is hampered by the fact that the  
adiabatic indices are different in these three simulations. The  
influence of the shape of the accretor, demonstrated by ZWN95, speaks  
against the transverse acoustic instability. These rather indirect  
arguments, if not conclusive, show at least that it is possible to  
reconcile existing simulations of 2D plane accretion in the framework  
of the advective-acoustic mechanism.

The stable 3D simulations can also be analyzed in the framework of the  
entropic-acoustic cycle. This cycle is stabilized by a weak shock, and  
destabilized by a small accretor size.
\par -Table~\ref{tableoverview} indicates that most axisymmetric  
simulations of an absorbing accretor are stable (SMT85, PSS89, SMA89),  
with the exception of KMS91 which shows an instability when $r_*/r_{\rm  
A}\le0.05$ and $\M\ge2.4$. The existence of a threshold for the  
accretor size and the minimum Mach number fits perfectly with the  
entropic-acoustic mechanism.

\par -In the axisymmetric simulations of FI98a with $v_\infty/c_{\rm  
light}=0.5$, the
Schwarzschild radius is too big to allow for a detached shock.  
According to Newtonian 3D simulations, the shock distance scales like a  
fraction (\eg typically $0.2-0.4$ in R94) of the accretion radius. A  
slower black hole would have a detached shock, and the  
entropic-acoustic instability could develop naturally if the  
temperature gradient is sufficient. The shock gets detached in the  
simulation of FI98a for $\gamma=5/3$, $v_\infty/c=0.15$, but the shock  
is then too weak ($\M_\infty=1.5$) to be destabilized. Indeed, the  
Newtonian simulations of Ruffert with $\M_\infty=1.4$ were also stable.  
A decisive test could be made by simulating an axisymmetric flow with  
$\gamma=5/3$, $\M_\infty=3$, and $v_\infty/c \le 0.1$.

\par -The apparent stability observed in the simulations of POM00, in  
particular for $\gamma=5/3$, cannot be explained by simple  
considerations about the accretor size and shock strength. This result  
may be attributed to the particular numerical method of local time step  
used by the authors. In Sect. 3 of POM00, the authors make the  
following statement:
  "It is important to note from the very begining that we seek steady  
state solution and generally do not perform time-accurate calculations  
(local time step inside each cell for the sake of computational  
efficiency)". This method might not be adequate to propagate high  
frequency acoustic waves across the subsonic region.

\subsection{Numerical artefacts in the simulations\label{artefacts}}

Numerical issues are numerous. Besides the damping effect of numerical  
viscosity (SPH and Eulerian codes were compared by BA94) and the  
possible axis effect in axisymmetric simulations (FTM87), more subtle  
effects can be understood through the advective-acoustic cycle.
This instability mechanism is physical, but may be artificially  
triggered or damped by numerical effects
such as the carbuncle phenomenon at the shock, the boundary condition  
at the surface of the accretor and the grid size in between.
The accuracy of both the advection of entropy/vorticity perturbations  
and the propagation of acoustic waves, between the shock and the sonic  
surface, is crucial for advective-acoustic instabilities.

\subsubsection{Carbuncle phenomenon at the shock}

POM00 drew attention to possible numerical instabilities at the shock  
in the BHL flow. In the region where the shock is parallel to the grid,  
the carbuncle instability (\eg Robinet \etal 2000) may favour the  
generation of vorticity and entropy perturbations, which in turn can  
feed an advective-acoustic cycle. Conversely, one should carefully  
check the effect of any numerical procedure designed to damp the  
carbuncle instability at the shock, since it may also damp the coupling  
between advected and acoustic perturbations.

\subsubsection{Grid resolution between the shock and the accretor}

\begin{figure}
\begin{center}
\includegraphics[width=\columnwidth]{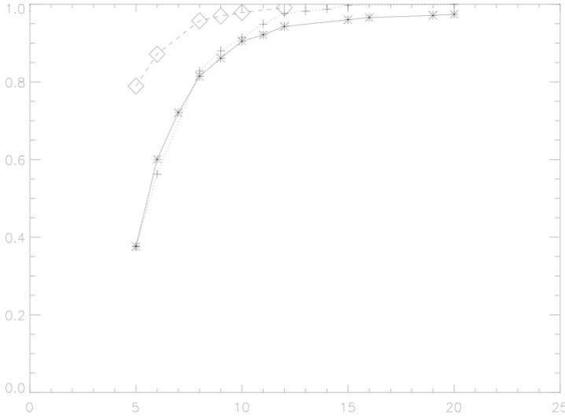}
\caption[]{Numerical damping of a sinusoidal perturbation of entropy  
(full line) or vorticity (dotted line) advected over one wavelength, in  
a direction parallel to the grid, as a function of the number of grid  
points per wavelength.
The flow is homogeneous with $\gamma=5/3$, and moves uniformly at Mach  
$0.2$.
The dashed line corresponds to the damping of an acoustic wave  
propagated over one wavelength. In the three cases, the wavevector of  
the perturbation is parallel to the flow.
}
\label{figgrid}
\end{center}
\end{figure}
Vorticity is not usually computed with as much accuracy as momentum or  
energy in classical numerical schemes such as those used for BHL  
simulations. The artificial generation of vorticity at the interface of  
nested grids could feed a vortical-acoustic cycle.\\
Conversely, insufficient numerical resolution is responsible for an  
artificial damping of vorticity and entropy waves. As an example,  
Fig.~\ref{figgrid} shows a measure of the damping of the amplitude of  
entropy and vorticity perturbations advected over one wavelength, in a  
direction parallel to the grid, as a function of the number of grid  
cells per wavelength. This test is performed on a Cartesian grid in 2D  
using the same PPM technique as in Ruffert (1994).
The correct advection of entropy and vorticity in this numerical  
simulation requires as much as 10-15 grid cells per wavelength.   
Perturbations with a wavelength shorter than 10 grid cells are  
significantly damped over one wavelength. The same test performed on  
acoustic waves propagating over one wavelength shows a smaller damping:  
5-10 grid cells are enough.
This is a strong constraint for the correct calculation of an  
advective-acoustic cycle at high frequency which involves perturbations  
with a wavelength comparable to the accretor size, when the sonic  
surface is attached to it. The grid size should thus be at least 15-20  
times smaller than the accretor size in order to properly describe the  
advective-acoustic coupling in the inner regions of the flow. This  
example illustrates the fact that the advective-acoustic instability  
can be impeded in a numerical simulation with insufficient resolution.  
Determining how these numbers depend on the numerical technique is  
beyond the scope of the present paper.

\subsubsection{Boundary condition on the accretor}

The size of the accretor is known to play an important role in the  
strength of the instability. One reason for this is that the sonic  
surface ahead of the accretor is generally attached to the accretor in  
numerical simulations. Boundary conditions at the surface of the  
accretor are therefore even more crucial, since they are in contact  
with a subsonic flow. The advection of entropy and vorticity  
perturbations through the boundary condition should be considered  
carefully in order to avoid an artificial acoustic feedback from the  
accretor.
For example, imposing a transverse velocity equal to zero in the  
perturbed Bondi flow would generate a spurious feedback $f^-$ from any  
non spherical entropy/vorticity perturbations $f^S$, and artificially  
excite a vortical-acoustic cycle. A calculation in Appendix~H of this  
condition at the boundary of a uniform, parallel flow shows that
\begin{eqnarray}
\left({f^-\over f^S}\right)_{\rm boundary}=-1\;\;{\rm if}\;\;l\ne0.
\end{eqnarray}
The artificial coupling due to inadequate boundary conditions can be as  
strong as the physical coupling ${\cal Q}_2$ expected from the  
temperature gradients in the flow.

\subsection{Physical or numerical instability?}

Many of the numerical artefacts discussed in Sect.~\ref{artefacts} were  
not considered by the authors of the existing simulations. Could new  
simulations of BHL accretion, corrected from numerical artefacts, be  
stable? The physical arguments of Sect.~\ref{adac} prove that the flow  
must be unstable, at least in the case $\gamma\sim 5/3$,  
$\M_\infty\ge3$, for a small enough accretor: in this case, the  
efficiency ${\cal Q}$ of the entropic-acoustic cycle diverges when the  
accretor size $r_*\to 0$. Even the leak of acoustic energy cannot  
diminish the efficiency ${\cal Q}$ by more than a finite geometrical  
factor $2-3$, at most. The argument is weaker for $\gamma<5/3$, because  
${\cal Q}$ depends strongly on the unknown shape of the sonic surface.

\section{Future numerical tests of the instability mechanism}

Future simulations should consider carefully the numerical artefacts  
listed in Sect.~\ref{artefacts}. The boundary condition at the surface  
of the accretor should be designed to absorb entropy and vorticity  
perturbations silently: this can be tested in a uniform flow.
An alternative would be to find a set of parameters such that the  
accretor is fully embedded inside the sonic surface. Since $\gamma=5/3$  
and $\gamma\sim1$ are ruled out by FR97 in 3D, an intermediate choice  
could be $\gamma=4/3$ with $\M_\infty=3$, and a small enough accretor.
The numerical issue of the absorbing boundary condition can also be  
solved naturally with general relativity, for any value of $\gamma$,  
since the flow is bound to be supersonic on the horizon of the black  
hole (FI98a,b).

The stable simulations analyzed in Sect.~\ref{stable} call for new  
simulations which would directly test the efficiency of the  
advective-acoustic mechanism in accretion flows where the shock is  
detached:
\par-all the axisymmetric, Newtonian simulations with $\gamma=5/3$  
should be unstable if $\M_\infty\ge 3$ and a small enough accretor  
($r_*/r_{\rm A}\sim 0.005$ seems to be enough according to KMS91). In  
particular, new axisymmetric, relativistic simulations similiar to  
FI98a should become unstable when considering a slower accretor with a  
moderate shock, such as $v_\infty/c_{\rm light}=0.1$, $\gamma=5/3$,  
$\M_\infty=3$. The axisymmetric simulations of POM00 should also be  
unstable if their numerical technique allows for the advective-acoustic  
cycle.
\par-2D planar simulation should be unstable through the  
entropic-acoustic cycle if the shock is detached: comparing a  
simulation with $\gamma\sim 3$ (see Appendix~A) and the classic  
flip-flop obtained for $\gamma\le 5/3$ could help to understand the  
respective influences of the entropic-acoustic cycle and the purely  
acoustic transverse instability.

In an unstable simulation, the advective-acoustic mechanism could be  
tested directly by measuring the incoming entropy/vorticity  
perturbations and the outgoing acoustic flux in the subsonic region of  
the flow, as measured by Blondin \etal (2004) in his simulations of  
spherical accretion. An indirect way of testing the instability  
mechanism is to measure the effect of the various physical parameters.  
The linear growth rate of the entropic-acoustic cycle should increase  
when the accretor size is decreased, decrease for a weak shock and  
saturate for strong shocks.
The present understanding of the 3D advective-acoustic instability when  
$\gamma\sim5/3$ implies that its strength for an accretor size $r_*\sim  
10^{-5}r_{\rm A}$ may be significantly under-estimated by the existing  
simulations ($r_*/r_{\rm A}>10^{-2}$). Tracing the power spectrum of  
the mass accretion rate (\eg R95), as a function of the numerical  
resolution in the range accessible to computers, could give a hint on  
the extrapolation to smaller accretor sizes.

In order to better understand the instability of isothermal flows, it  
would be interesting to try
to discriminate between a transverse instability (purely acoustic) and  
a vortical-acoustic cycle.
The accretor size and boundary condition should play a more important  
role in a vortical-acoustic cycle than in a transverse instability.

\section{Conclusion}

The physical mechanisms underlying the longitudinal and transverse  
instabilities of the accretion line have been clarified. The  
destabilizing factor for the longitudinal instability is the density  
dependence of the external force per unit mass. The transverse  
instability is ruled by the restoring force proportional to the local  
inclination of the accretion line. The analogy with the instability of  
a flag has been outlined. \\
The overstable amplification of longitudinal high frequency  
perturbations of the accretion line found by Cowie (1977) is greatly  
affected by pressure forces. Density perturbations are propagated as  
acoustic waves. Those propagating outwards are damped, whereas those  
propagating inwards are transiently amplified. The analogy with the  
instability of radiation driven winds has been drawn. \\
The transverse instability of the accretion line is limited if the  
finite width of the shock cone is taken into account. A feedback  
process is necessary to explain the global flip-flop instability  
observed in planar 2D simulations. It seems possible that the mechanism  
of the transverse instability, modified by the propagation of acoustic  
waves within the subsonic region of the flow, leads to a global  
unstable mode. However, the influence of the shape of the accretor,  
demonstrated by ZWN95, may be a hint against this explanation of the  
instability. \\
For the first time, the advective-acoustic instabilitiy has been  
analyzed in the context of BHL accretion. This analysis is based on the  
extrapolation of the properties established in spherically symmetric  
flows (F01, F02). For this reason, the relevance of the  
advective-acoustic instabilities in BHL accretion is convincingly  
demonstrated only when the shock is detached from the accretor. The  
difference of geometry precludes an accurate prediction of the  
instability threshold in BHL accretion, especially since it is very  
sensitive to the size and shape of the sonic surface.
Nevertheless, the analysis is predictive enough to assess that the  
advective-acoustic mechanism
\par - must be stable in the limit of a weak shock,
\par - must be unstable for 3D accretion flows with $\gamma\sim 5/3$, a  
reasonable shock strength $\M_\infty\ge 3$ and a small enough accretor  
size.\\
Numerical artefacts have also been discussed, specifically the grid  
size and the boundary conditions. We comment on how these artefacts  
must be circumvented to
produce reliable numerical simulations of BHL accretion. \\
Surprisingly, it seems that all the existing numerical simulations can  
be reconciled in the framework of the advective-acoustic instability,  
with no striking contradiction, even when the shock is attached to the  
accretor. Several numerical tests of these ideas have been proposed.\\

Besides new numerical simulations, future analytic work ought to  
describe in more detail both the transverse instability modified by  
pressure forces, and the efficiency of the advective-acoustic cycle  
when the shock is attached. This could help understand why isothermal  
accretion is so much more unstable in 2D than in 3D.

\begin{acknowledgements}
The authors are grateful to Egide and the British Council for their  
exchange programme. The anonymous referee is thanked for his  
constructive comments.
\end{acknowledgements}

\appendix

\section{BHL stationary flow in 2D and 3D}

\subsection{Adiabatic index and pressure forces}

The sonic radius in radial Bondi accretion, deduced from the Bernoulli  
equation and the conservation of mass flux, is different in 2D and 3D:
\begin{eqnarray}
r_{\rm son}&=&{3-\gamma\over 2}{GM\over c_\infty^2}\;\;{\rm  
in\;2D},\label{rson2D}\\
r_{\rm son}&=&{5-3\gamma\over4}{GM\over c_\infty^2}\;\;{\rm  
in\;3D}.\label{rson3D}
\end{eqnarray}
The existence of a sonic radius in radial accretion on a point like  
accretor requires $\gamma\le5/3$ in 3D, whereas plane supersonic  
accretion is possible up to $\gamma\le3$. This illustrates the fact  
that the 3D convergence of flow lines produces stronger pressure  
gradients than in 2D. These pressure gradients act against gravity.
Extending to plane flows the argument used in FR97 for 3D flows, the  
sonic surface of a stationary BHL accretion must intersect the sphere  
of radius $r_0$ deduced from
Eqs.~(\ref{rson2D}-\ref{rson3D}), with
\begin{eqnarray}
r_0&\equiv&{r_{\rm son}\over 1+{\gamma-1\over2}\M_\infty^2}.
\end{eqnarray}
This suggests that the shock should be detached in planar accretion  
with $\gamma$ close to 3. SMA89 considered plane accretion with  
$\gamma=2$ but the shock was still not detached. The ratio $r_0/r_{\rm  
A}$ is a function of $\gamma,\M_\infty$:
\begin{eqnarray}
{r_0\over r_{\rm A}}&=&{3-\gamma\over 4}
{\M_\infty^2\over 1+{\gamma-1\over2}\M_\infty^2}\;\;{\rm  
in\;2D},\label{rson2Db}\\
{r_0\over r_{\rm A}}&=&{5-3\gamma\over8}
{\M_\infty^2\over 1+{\gamma-1\over2}\M_\infty^2}\;\;{\rm  
in\;3D}.\label{rson3Db}
\end{eqnarray}
As already noted in FR97, the sonic surface must also be attached to  
the accretor if $r_0/r_{\rm A}\ge1$, because the sonic surface cannot  
extending beyond the distance $\sim r_A$ of the stagnation point. This  
concerns in particular isothermal flows with a strong shock, for which  
$r_0/r_{\rm A}\propto\M_\infty^2$. A critical index $\gamma_{\rm crit}$  
must therefore exist, below which the shock is always attached to the  
accretor, whatever its size.

\subsection{Accretion line}

The mass flux per unit of length, along the accretion line, is constant  
in 3D whereas it varies like $r^{-{1\over2}}$ for a planar flow (Soker  
1990). By integrating Eq.~(\ref{dzH}) using Table~\ref{tableacline},
\begin{eqnarray}
\rho v&=&2(r^{1\over 2}-\alpha^{1\over 2})\;\;{\rm  
in\;2D},\label{rho2D}\\
\rho v&=&r-\alpha\;\;{\rm in\;3D},\label{rho3D}
\end{eqnarray}
where $\alpha$ is the distance of the stagnation point.
The asymptotic velocity within the accretion line, deduced from  
Eq.~(\ref{dyF}) for $r\gg\alpha$, is different in 2D and 3D:
\begin{eqnarray}
v&\sim&1-{\lambda\over r^{1\over 2}}\;\;{\rm in\;2D},\label{v2D}\\
v&\sim&1-{\log r\over r}\;\;{\rm in\;3D},\label{v3D}
\end{eqnarray}
where $\lambda$ is a constant.

\section{Physical cause of the longitudinal instability of the  
accretion line}

The growth rate computed by Cowie (1977) in 3D is the imaginary part of  
the complex frequency $\omega$:
\begin{eqnarray}
(\omega-kv)^2&=& ik\left\lbrack v{\d v\over\d r}+{1\over 2r^2  
}\right\rbrack.\label{omecowie}
\end{eqnarray}
The growth rate in 2D is given by strictly the same formula, although  
the accretion terms involve geometrical factors $1/r^{1\over2}$ in  
Table~\ref{tableacline}. The absence of such factors in  
Eq.~(\ref{omecowie}) led Soker (1990) to conclude that this instability  
is independant of accretion. This argument is insufficient, since  
Eq.~(\ref{omecowie}) can be rewritten using the stationary flow  
equation Eqs. (2.7a-2.7b) of Soker (1990), as follows:
\begin{eqnarray}
(\omega-kv)^2&=&
{ik\over2}{1-v\over \rho r^{1\over2}}\;{\rm in\;2D},\\
&=&{ik\over2}{1-v\over \rho }\;{\rm in\;3D}.
\end{eqnarray}
The geometrical factor $r^{1\over2}$ is then clearly apparent.
A closer look at the equations shows that the longitudinal instability  
is due to the density dependance of the force acting on the accretion  
line.
Using the same normalizations of density, velocity and distances as in  
Soker (1990), the time dependent equations correspond to  
Eqs.~(\ref{dzH}-\ref{dyF}).\\
A linearization of Eqs.~(\ref{dzH}-\ref{dyF}) gives the differential  
equation satisfied by a perturbation of the mass flux $h\equiv  
\rho\delta v+v\delta\rho$:
\begin{eqnarray}
v^2{\p^2h\over \p r^2}-{\p h\over\p r}\left\lbrace
2i\omega v-\rho{\p\over\p r}{v^2\over \rho}
+v{\p F\over\p v}-\rho{\p F\over\p \rho}\right\rbrace
\nonumber\\
-i\omega h\left\lbrace \rho{\p\over\p r}{v\over \rho}-i\omega
-{\p F\over\p v}\right\rbrace=0.\label{singlediff}
\end{eqnarray}
The solution is written in Eq.~(\ref{wkbh}) at high frequency $\omega$  
using the WKB approximation. Contrary to the conclusions of Soker  
(1990), acceleration within the accretion line is not crucial for this  
instability. The simplest flow in which a similar instability occurs  
would be a flow with uniform density $\rho_0$ and velocity $v_0$,  
subject to a force depending linearly on density: $F\equiv  
a(\rho-\rho_0)$ and without any mass accretion ($H\equiv 0$). The  
stationary flow being uniform, the uniform velocity could also be taken  
to be equal to zero owing to a simple change of reference frame.
The evolution of perturbations can then be calculated precisely. The  
frequency $\omega$ and the wavevector $k$ are related through the  
following dispersion relation:
\begin{eqnarray}
(\omega-kv_0)^2=ik\rho_0{\p F\over\p \rho}
\end{eqnarray}
The growth rate corresponds to the imaginary part $\omega_i$:
\begin{eqnarray}
\omega_i={a\rho_0\over 2(\omega_r -kv_0)}
\end{eqnarray}
The amplification of perturbation is thus exponential in the direction  
of the external force for a positive enhancement of density.
In the accretion line model, this amplification is weaker due to the  
weak density dependence of the external force. Using the asymptotic  
behaviour $v\propto -1/r^{1\over2}$ close to the accretor, and  
Eq.~(\ref{v3D}) in 3D flows leads to  
Eqs.~(\ref{ampliout3}-\ref{ampliin3}). By contrast, the same  
calculation using Eq.~(\ref{v2D}) in 2D flows leads to:
\begin{eqnarray}
h(r)\propto r^{3\over8}\exp\left\lbrack i\omega r
\pm2(1+i)\omega^{1\over2}\lambda^{1\over2}r^{1\over4}\right\rbrack,\nonumber\\
{\rm for}\;\;r\gg\alpha,\label{ampliout2}\\
h(r)\propto r^{3\over4}\exp
\left\lbrack-{2\over3}i\omega r^{3\over 2}
\pm{1-i\over  
2\alpha^{1\over2}}\omega^{1\over2}r\right\rbrack,\nonumber\\
{\rm for}\;\;{1\over\omega^{1\over 2}}\ll r\ll\alpha.\label{ampliin2}
\end{eqnarray}
The difference of asymptotic behaviors between the 2D and 3D cases is  
not significant.\\

\section{Effect of pressure forces on the longitudinal instability}

The linearization of the perturbed flow leads to define the  
perturbation $f$ of the Bernoulliu constant as follows:
\begin{eqnarray}
f&\equiv& v\delta v+c^2{\delta\rho\over \rho},
\end{eqnarray}
in order to obtain a simple second order differential system satisfied  
by $(f,g)$:
\begin{eqnarray}
(c^2-v^2){\p h\over\p r}&=&i\omega \left( \rho f-v h \right),\\
(c^2-v^2){\p f\over \p r}&=&\left\lbrace -i\omega v -v{\p F\over\p  
v}+\rho{\p F\over\p \rho }
\right\rbrace f\nonumber\\
&&+\left\lbrace \left(i\omega +{\p F\over\p v}\right)c^2-v\rho{\p  
F\over\p \rho }\right\rbrace {h\over \rho}.
\end{eqnarray}
A single differential equation is obtained:
\begin{eqnarray}
(c^2-v^2){\p^2h\over \p r^2}+{\p h\over\p r}\left\lbrace
2i\omega v+\rho{\p\over\p r}\left({c^2-v^2\over \rho}\right)
\right.\nonumber\\
\left.+v{\p F\over\p v}-\rho{\p F\over\p \rho}\right\rbrace
+i\omega h\left\lbrace \rho{\p\over\p r}{v\over \rho}-i\omega
-{\p F\over\p v}\right\rbrace=0.\label{singlediffpress}
\end{eqnarray}
Before looking for approximate solutions to this equation, the effect  
of pressure forces can be easily incorporated in the simple toy model  
used in Appendix~B, with uniform density and velocity:
\begin{eqnarray}
\omega={i\over2}{\p F\over\p v}+kv\pm \left\lbrack
k^2c^2+ik\rho{\p F\over\p\rho}-{1\over 4}\left({\p F\over\p v}\right)^2
\right\rbrack^{1\over2}
\end{eqnarray}
The problem is formally identical to that idealized by  Mestel, Moore  
\& Perry (1976) and Mathews (1976), for radiation driven winds, where  
$F\equiv A\rho-g$. At high frequency:
\begin{eqnarray}
\omega\sim k(v\pm c)+{i\over2}\left({\p F\over\p v}
\pm{\rho\over c}{\p F\over\p\rho}\right).
\end{eqnarray}
The stability of high frequency acoustic waves depends on the sign of  
the quantity between parenthesis. In a non uniform flow, a similar  
conclusion can be reached at high frequency through a WKB analysis,  
leading to Eq.~(\ref{freqwkb}).

\section{Transverse instability of the accretion line}

The differential equation satisfied by $\g$ is:
\begin{eqnarray}
{\p^2 \g\over \p r^2}+{\p \g\over\p r}\left\lbrace{\p\log r^2 v\over\p  
r}-{2i\omega\over v}
+{1\over r^{1\over 2}v\rho }\left(1+{1\over  
v}\right)\right\rbrace\nonumber\\
-{\g\over r\rho v^2}\left\lbrace r\rho \omega^2+2i\omega \rho v+i\omega  
r^{1\over2}-{3\over 2r^{1\over 2}}
\right\rbrace=0.\label{eqdifg}
\end{eqnarray}
The WKB approximation is thus:
\begin{eqnarray}
\g\sim {(\rho v)^{1\over 4}\over r^{7\over8}}\exp\int\left({i\omega  
\over v}
-{1\over 2r^{1\over 2}v\rho }\right)\d r\nonumber\\
\exp \pm {1+i\over 2^{1\over 2}}\omega^{1\over 2}\int{\d r\over  
r^{1\over4}\rho^{1\over 2}v^{3\over 2}},
\;\;{\rm for}\;\;\omega\gg{3v\over 2r}.\label{transversewkb}
\end{eqnarray}
The second order differential equation in the general case  
corresponding to Eq.~(\ref{geneflop}) is:
\begin{eqnarray}
v^2{\p^2\g\over\p r^2}+{\p \g\over\p r}\left(
{v\over r}{\p rv\over\p r}-2i{\omega v}-B-Cv\right)\nonumber\\
+\g\left(-i\omega {v\over r}-{A+B\over r}+i\omega C-\omega^2  
\right)=0.\label{diflag}
\end{eqnarray}

\section{Proof that $\delta K\equiv0$ in the shocked Bondi flow}

Let us recall the differential system satisfied by the perturbations  
$f,g$ defined in F01 (Eqs.~(B18-B19)):
\begin{eqnarray}
v{\p f\over\p r}+{i\omega \M^2f\over 1-\M^2}
&=& {i\omega v^2 g\over 1-\M^2} + i\omega c^2 {\delta
S\over\gamma},
\label{dfdr}\\
v{\p g\over\p r}+{i\omega \M^2g\over 1-\M^2}
&=& {i\omega f\over c^2(1-\M^2)} - {iL^{2}\over \omega r^2}f+
{i\delta K\over r^2\omega },\label{dgdr}
\end{eqnarray}
where the constant $\delta K$ and the function $\mu(r,\omega,l)$ are  
defined by:
\begin{eqnarray}
\delta K&\equiv& r^2v\cdot (\nabla\times \delta w)+l(l+1)c^2
{\delta S\over\gamma},\label{defK}\\
\mu^{2}&\equiv&1 - {l(l+1)\over \omega^2r^2}(c^{2}-v^{2}).\label{defmu}
\end{eqnarray}
Consider a spherical adiabatic shock with incident Mach number $\M_1$  
in the
radial direction. Let this shock be perturbed by a sound wave with
frequency $\omega$ propagating against the flow in the subsonic region,
producing a displacement $\Delta\zeta(\theta,\varphi)$ and a
perturbation $\Delta v(\theta,\varphi)$ of the radial velocity of the  
shock.
Using the index "1" before the shock, and "2" after it, the  
conservation of
mass flux and energy across the shock can be written as follows:
\begin{eqnarray}
\rho_{1}(v_{1}-\Delta v)&=&\rho_{2}(v_{2}+\delta v_{2}-\Delta v),\\
{(v_{1}-\Delta v)^{2}\over2}+{c_{1}^{2}\over\gamma-1}&=&
{(v_{2}+\delta v_{2}-\Delta v)^{2}\over2}
+{(c_{2}+\delta c_{2})^{2}\over\gamma-1},
\end{eqnarray}
where quantities are measured at the position $\rsh+\Delta\zeta$.  
Keeping
the first order terms, and using the defnition of $f,g$, together with
the entropy equation, we obtain:
\begin{eqnarray}
f&=&(v_{2}-v_{1})\Delta v,\label{boundf}\\
g&=&\left({1\over v_{2}}-{1\over v_{1}}\right)\Delta v+\delta  
S.\label{boundg}
\end{eqnarray}
A third equation relating $\delta S$ to $\Delta v,\Delta\zeta$ could be  
deduced
using the conservation of impulsion, in the spirit
of Nakayama (1994). A more direct derivation can be obtained by
noting that entropy is conserved before and after the shock, and
that the entropy jump across the shock depends only on the local
value of the incident Mach number $\M_{1}'$ in the frame of the shock
(Eq.~(\ref{zetav})):
\begin{eqnarray}
\M_{1}'(\rsh+\Delta\zeta)&=&\M_{1}(\rsh)+{\Delta v\over c_{1}}+
\Delta\zeta{\p\M_{1}\over\p r},\label{zetav}\\
&=&\M_{1}(\rsh)+\left(1+{i\eta v_2\over\omega r}\right)
{\Delta v\over c_{1}},\\
\eta&\equiv&{\p\log\M\over\p\log r}
\end{eqnarray}
Using the Rankine-Hugoniot jump conditions, we obtain
\begin{equation}
{\delta S}=-{4\gamma(\M_1^2-1)^2\over
(\gamma+1)^{2}\M_{1}^{2}}{c_{1}^{2}\over c_{2}^{2}}
\left(1+{i\eta v_2\over\omega r}\right){\Delta v\over v_1}.\label{dsv1}
\end{equation}
Equations (\ref{boundf}-\ref{boundg}) thus become:
\begin{eqnarray}
f&=&-{2c_{1}^{2}\over\gamma+1}(\M_{1}^{2}-1){\Delta v\over
v_{1}},\label{boundf1}\\
g&=&{2\over\gamma+1}{c_{1}^{2}\over c_{2}^{2}}
{\M_{1}^{2}-1\over\M_{1}^{2}}\left\lbrack
1-{2\gamma\over\gamma+1}(\M_{1}^{2}-1){i\eta v_2\over\omega r}
\right\rbrack{\Delta v\over v_{1}}.\label{boundg1}
\end{eqnarray}
The perturbation of non radial velocity
$\delta v_\theta,\delta v_\varphi$ immediately after the shock is  
deduced
from the
conservation of the tangential component of the velocity across the
shock, in the spirit of Landau \& Lifschitz (1987).
\begin{eqnarray}
\delta v_\theta&=&{v_{1}-v_{2}\over r}{\p \Delta \zeta\over \p \theta},
\label{dvt}\\
\delta v_\varphi&=&{v_{1}-v_{2}\over r\sin\theta}
{\p \Delta \zeta\over \p \varphi}\label{dvp}.
\end{eqnarray}
The vorticity immediately after the shock is deduced from the non
radial component of the linearized Euler equations (B.11) and (B.12) of  
F01,
together with Eq.~(\ref{boundf}) and Eqs~(\ref{dvt}-\ref{dvp}):
\begin{eqnarray}
w_{r}&=&0,\label{wr}\\
w_{\theta}&=&-{c^{2}\over rv\sin\theta}{\p\over\p\varphi}
{\delta S\over\gamma}\label{wt},\\
w_{\varphi}&=&{c^{2}\over rv}{\p\over\p\theta}
{\delta S\over\gamma}\label{wp}.
\end{eqnarray}
Immediately after the shock, the constant $\delta K$ defined by
Eq.~(\ref{defK}) is computed using Eqs.~(\ref{wt}) and (\ref{wp}):
\begin{equation}
\delta K=0.
\end{equation}
Since $\delta K$ is conserved through the Bondi flow, $\delta K$ is
uniformly equal to zero. As a consequence, using the integrated  
expression of the vorticity (Eqs.~(B5) to (B7) in F01), the vorticity  
perturbation is described by Eqs.~(\ref{wr}) to (\ref{wp}) throughout  
the flow.

\section{Advective-acoustic coupling at the shock}

The perturbations $f,g$ after the shock are decomposed as follows:
\begin{eqnarray}
f&=&f^{-}+f^{+}+f^S,\label{decompf}\\
g&=&g^{-}+g^{+}+g^S,\label{decompg}
\end{eqnarray}
where $f^S,g^S$ correspond to the entropy/vorticity wave
associated to the entropy perturbation $\delta S$ with $\delta K=0$, and
$f^{\pm},g^{\pm}$ correspond to the purely acoustic waves propagating
in the direction of the flow (index $+$) or against the flow
(index $-$).
Neglecting the coupling between the entropic and acoustic waves in
the vicinity of the shock, the entropy wave $f^S,g^S$ is advected
at the velocity of the fluid:
\begin{eqnarray}
{\p f^S\over\p r}&\sim& {i\omega \over v}f^S,\\
{\p g^S\over\p r}&\sim& {i\omega \over v}g^S.
\end{eqnarray}
Replacing these derivatives in Eqs.~(\ref{dfdr}-\ref{dgdr}), we obtain:
\begin{eqnarray}
f^S&\sim&{1-\M_2^2\over 1-\mu^2\M_2^2}
c^{2}{\delta S\over\gamma},\label{fe}\\
g^S&\sim&{\mu^2\over c^2}f^S.\label{ge}
\end{eqnarray}
Acoustic waves are described by Eqs.~(\ref{dfdr}-\ref{dgdr}) in the
absence of entropy perturbations, \ie when $\delta S=0$.
Using the WKB approximation of F01, the radial
derivative of $f^\pm$ is approximated by:
\begin{equation}
{\p f^\pm\over\p r}\sim{i\omega\over c}{\M \mp  
\mu\over1-\M^{2}}f_\pm.\label{dfwkb}
\end{equation}
$g^{\pm}$ is deduced from Eqs.~(\ref{dfdr}) and (\ref{dfwkb}):
\begin{equation}
g^\pm\sim\pm {\mu\over\M c^{2}}f_{\pm}.\label{gwkb}
\end{equation}
The linear system (\ref{decompf}-\ref{decompg}) can be transformed  
using Eqs.(\ref{fe}-\ref{ge}) and (\ref{gwkb}),  in order to express  
the acoustic perturbations $f^\pm$:
  \begin{eqnarray}
f^{\pm}={1\over 2 }\left\lbrack f \pm {{\cal M}\over\mu}c^2g-
(1\pm\mu\M)f^S
  \right\rbrack.\label{fpmgen}
\end{eqnarray}
Using this equation immediately after the shock, with Eqs.~(\ref{dsv1})  
to (\ref{boundg1}):
\begin{equation}
f^{\pm}\sim\mp{\gamma+1\over4\gamma}{c^{2}\delta S\over 1-\M_{1}^{-2}}
{\M_{2}\over\mu}\left({\mu^{2}\mp  
2\M_{2}\mu+\M_{1}^{-2}\over1\mp\mu\M_{2}}
\right).\label{fpm2}
\end{equation}
 From Eqs.~(\ref{fe}) and (\ref{fpm2}),
\begin{eqnarray}
{f^S\over f^-}={4\over\gamma+1}{\mu\over\M_2}
{(1-\M_2^2)(1-\M_1^{-2})\over
(1-\mu\M_2)
(\mu^2+2\mu\M_2+\M_1^{-2})}.
\end{eqnarray}
At high frequency such that $\mu\sim 1$,
\begin{eqnarray}
{f^S\over f^-}&=&{4\over\gamma+1}
{1+\M_2\over1+2\M_2+\M_1^{-2}}
{1-\M_1^{-2}\over\M_2},\\
&\propto&{1-\M_2\over\M_2}.
\end{eqnarray}

\section{Entropic-acoustic coupling in 3D for $\gamma\sim5/3$}

The value of $|{\cal Q}|$ deduced from F01
for $\gamma$ close to $5/3$ increases with frequency like
$\omega_r^{1/3}$ (Eqs.~28-29 of F01), up to a maximum
reached near the cut-off frequency. This behaviour can be understood
in the framework of FT00, which argued that the efficiency of
the entropic-acoustic coupling is related to the increase of
enthalpy between the shock and the sonic point. A slight correction,  
however,
should be made. Rather than the naive guess $|{\cal Q}|^{2}\propto
c_{\rm son}^{2}/c_{\rm sh}^{2}$ of Eq.~(23) in FT00, one must
take into account the fact that the coupling of entropy perturbations to
acoustic waves must occur before the sonic radius in order to allow
significant outgoing acoustic flux. The effective radius of coupling
$r_{\rm eff}$ was computed analytically in Appendix~E of F01
(Eqs.~E6 and E13):
\begin{eqnarray}
r_{\rm eff}&\propto & \omega_r^{-{2\over3}},\\
|{\cal Q}|^{2}&\propto & {c^{2}(r_{\rm eff})\over c^{2}(r_{\rm sh})}
\propto  \omega_r^{{2\over3}}.
\end{eqnarray}
The efficiency ${\cal Q}^{2}$ deduced from the analytical
calculations of F01 indeed scales like the ratio of enthalpies between
the shock and the effective point of coupling $r_{\rm eff}$.
The radius $r_{\rm eff}$ coincides with the wavelength $v/\omega_{\rm  
cut}$ of an entropy
perturbation with a frequency $\omega_r$ close to $\omega_{\rm cut}$:  
the
enthalpy effectively "seen" by this perturbation is not the enthalpy
at the sonic radius but rather at the effective radius $r_{\rm eff}$.
 From this point of view, using the same notations as in F01, $|{\cal  
Q}_{S}|_{l=1}\gg |{\cal Q}_{S}|_{l=0}$
is a natural consequence of $\omcutu\gg\omcutz$: non radial  
perturbations
effectively "see" regions of higher enthalpy.
The most efficient entropic-acoustic coupling is reached at frequencies
close to the refraction cut-off, where $\mu\sim1$.

\section{Artificial acoustic feedback from the boundary condition on  
the accretor}

The perturbed velocity in the direction perpendicular to the flow is  
deduced from the linearized Euler equations, using the vorticity given  
by Eqs.~(\ref{wt}-\ref{wp}):
\begin{eqnarray}
\delta v_\theta&=&{1\over i\omega r}{\p f\over\p\theta},\\
\delta v_\varphi&=&{1\over i\omega r\sin\theta}{\p f\over\p\varphi}.
\end{eqnarray}
A combination of these equations involves the eigenvalues of the  
Laplacian in spherical coordinates:
\begin{eqnarray}
l(l+1)f
   &=&-{i\omega r\over {\rm sin}\theta}\left\lbrack{
\partial\over
\partial\theta}({\rm sin}\theta \delta v_{\theta})+{
\partial\over
\partial\phi}\delta v_{\phi}\right\rbrack.
\end{eqnarray}
Imposing $\delta v_\perp=0$ on the boundary consequently requires $f=0$  
there, if the perturbation is not spherically symmetric ($l\ne 0$).
The acoustic feedback $f^-$ associated to the entropy/vorticity  
perturbation $f^S$ passing through the boundary condition $\delta  
v_\perp=0$, is deduced from the decomposition (\ref{decompf}), with  
$f^+=0$ and $f=0$:
\begin{eqnarray}
\left({f^-\over f^S}\right)_{\rm boundary}&=&-1\;\;{\rm if}\;\; l\ne0.
\end{eqnarray}

\end{document}